\documentclass[aps,10pt,notitlepage,twocolumn,nofootinbib,superscriptaddress]{revtex4-2}

	\usepackage{graphicx,color}
	\usepackage{amsmath}
	\usepackage{amssymb}
	\usepackage{hyperref}
	\usepackage[utf8]{inputenc}
	\usepackage[english]{babel}
	\usepackage{epsfig}
	\usepackage{subfigure}
	\usepackage{wasysym}
	\usepackage{color,xcolor}
	\usepackage{amsmath}
	\usepackage{bm}
	\usepackage{epsfig}
	\usepackage{amsfonts}
	\usepackage{dcolumn}
	\usepackage{float}
	\usepackage{booktabs}
	\usepackage{relsize}
	
\begin{document}

\title{Quantum-inspired wormholes from String T-Duality}


	\author{Francisco S. N. Lobo} \email{fslobo@ciencias.ulisboa.pt}
\affiliation{Institute of Astrophysics and Space Sciences, Faculty of Sciences, University of Lisbon, Building C8, Campo Grande, P-1749-016 Lisbon, Portugal}
\affiliation{Department of Physics, Faculty of Sciences, University of Lisbon, Building C8, Campo Grande, P-1749-016 Lisbon, Portugal}

    \author{Manuel E. Rodrigues} \email{esialg@gmail.com}
\affiliation{Faculty of Physics, Graduate Program in Physics, Federal University of Pará, 66075-110, Belém, Pará, Brazil}
\affiliation{Faculty of Exact Sciences and Technology, Federal University of Pará, Abaetetuba University Campus, 68440-000, Abaetetuba, Pará, Brazil}

\begin{abstract}

The intrinsic non-perturbative features of string-theoretic corrections, particularly those arising from T-duality, have been shown to naturally introduce an effective ultraviolet (UV) cutoff into the gravitational framework. This cutoff, often referred to as the zero-point length in the context of path integral duality, acts as a fundamental minimal length scale that regulates short-distance divergences. Using the established correspondence between T-duality and path integral duality, it has been shown that the static Newtonian potential becomes regularized at small distances. Building upon this regular behavior, we proceed to construct self-consistent, spherically symmetric, and electrically neutral wormhole solutions, which remain free of curvature singularities and embodies the effects of this duality-induced UV completion.
We explore wormhole configurations with the aim of minimizing violations of the null energy condition (NEC). In fact, solutions with a constant redshift function or specific shape functions generally exhibit NEC violations throughout the entire spacetime. To address this, we explore two distinct thin shell constructions: (i) wormholes formed by matching an interior wormhole geometry to an exterior Schwarzschild vacuum spacetime, thereby confining exotic matter to a localized region; and (ii) standard thin-shell wormholes generated by gluing two identical black hole spacetimes across a timelike hypersurface situated outside their event horizons. In both cases, the NEC violations are minimised and restricted to a finite region, improving the physical plausibility and traversability of the resulting configurations.

\end{abstract}

\date{\today}
\maketitle
\def\HMS{{\scriptscriptstyle{\rm HMS}}}

\section{Introduction}

One of the enduring challenges in formulating a theory of quantum gravity is the emergence of a fundamental length scale below which the classical notion of spacetime ceases to be meaningful. Early investigations into this idea showed that incorporating such a minimal length—typically expected to be of the order of the Planck length, $L_P$—modifies the structure of the spacetime interval and the corresponding quantum propagators. In particular, Padmanabhan demonstrated that requiring invariance of the path integral amplitude under the duality transformation $ds \rightarrow L_P^2/ds$ leads to a modified propagator in which the spacetime interval $(x-y)^2$ is replaced by $(x-y)^2 + L_P^2$, suggesting the notion of a ``zero-point length'' \cite{Padmanabhan:1996ap}. This modification serves as a natural ultraviolet (UV) regulator and encodes quantum gravitational corrections into low-energy physics.

These insights have been further substantiated in the context of string theory, where T-duality—an inherent feature of extended objects—plays a key role in inducing similar modifications in propagators. This supports the duality-based path integral approach and strengthens the interpretation of the zero-point length as an effective description of stringy corrections at low energies \cite{Smailagic:2003hm}. Moreover, the work of Fontanini et al.~\cite{Fontanini:2005ik} demonstrated that the zero-point length in four-dimensional spacetime can be consistently understood as a residual imprint of compact extra dimensions. The magnitude of this length is found to be directly linked to the string length scale through T-duality, within the context of a modified Kaluza-Klein framework. This connection provides a deeper understanding of how string theoretical effects manifest at macroscopic scales, particularly in gravitational systems.

A compelling application of the zero-point length framework emerges in black hole spacetimes modified by string T-duality. In Ref.~\cite{Nicolini:2019irw}, it was shown that stringy corrections associated with T-duality introduce a fundamental length scale acting as an ultraviolet (UV) regulator, effectively embedding non-perturbative quantum gravity effects into low-energy gravitational dynamics. By exploiting the correspondence between T-duality and path integral duality, the authors derived a regularized static Newtonian potential incorporating a minimal length identified with the zero-point length.
The resulting potential can be interpreted as arising from a smeared matter distribution, in contrast to the singular profiles associated with conventional pointlike sources. 
The corresponding energy density of this modified matter configuration is determined via the Poisson equation~\cite{Nicolini:2019irw}
\begin{equation}
	\rho(r)
	= \frac{1}{4\pi}\triangle V(r)
	= \frac{3 l_0^2 M}{4 \pi {\left(r^2 +l_0^2\right)}^{5/2}}\,,
	\label{eq:rho0}
\end{equation}
where $l_0$ denotes the zero-point length of spacetime. This potential was then used to construct a consistent, spherically symmetric and electrically neutral black hole metric, which is completely regular and structurally resembles the Bardeen solution \cite{2767662}, differing only in the specific UV cutoff employed. On the thermodynamic side, the modified black hole exhibits a maximum Hawking temperature followed by a cooling phase, culminating in a stable remnant. These results support the idea of universality in quantum-corrected black holes and demonstrate how T-duality can naturally resolve curvature singularities.

This approach was further generalized in Ref.~\cite{Gaete:2022ukm} to encompass charged and rotating black hole solutions within the same T-duality framework. The resulting spacetimes remain regular and exhibit structural similarities to well-known solutions derived from non-linear electrodynamics, such as those proposed in Refs.~\cite{Ayon-Beato:1998hmi,Ayon-Beato:1999kuh}, particularly when the T-duality length scale is identified with the electric charge. In this construction, the use of Padmanabhan’s propagator \cite{Padmanabhan:1996ap} --- modified by the zero-point length and inherently encoding T-duality effects --- plays a central role in generating the static potential and ensuring regularity at short distances. These results not only reinforce the robustness of T-duality as a mechanism for singularity resolution but also emphasize its phenomenological relevance in constructing physically viable, quantum-corrected black hole geometries that smoothly extend classical solutions.

Further implications of the T-duality framework were explored in Ref.~\cite{Jusufi:2023dix}, where a pair of regular black holes incorporating T-duality corrections was used to construct an Einstein-Rosen (ER) bridge with a throat radius proportional to the Planck length. This setup offers a geometric model for quantum entanglement, lending support to the ${\rm ER}={\rm EPR}$ conjecture \cite{Maldacena:2013xja}, which links spacetime geometry to quantum correlations. For extremal configurations, the horizon area matches the Bekenstein bound and the wormhole mass approaches the Planck scale, suggesting a realization of gravitational self-completeness. In the sub-Planckian regime, horizonless wormholes emerge with localized regions of negative energy near the throat, possibly arising from quantum fluctuations or Casimir effects. This highlights the role of T-duality in both resolving singularities and probing the quantum structure of spacetime.

Building upon the regularized behavior of the gravitational potential at short distances, arising from duality-inspired ultraviolet (UV) completions such as those induced by T-duality, as mentioned above, we proceed to construct self-consistent, spherically symmetric, and electrically neutral wormhole solutions. These geometries remain free of curvature singularities and incorporate the fundamental minimal length scale associated with the duality framework, effectively capturing non-perturbative string-theoretic corrections in the low-energy regime.
A central feature in the theoretical construction of traversable wormholes is the violation of the null energy condition (NEC), which implies the existence of exotic matter~\cite{Morris:1988cz}. This exotic matter is typically required to support the wormhole throat and prevent gravitational collapse. In this work, we aim to minimize and spatially confine such violations. To achieve this, we explore two distinct classes of thin shell wormhole configurations, both designed to localize NEC violations within finite, well-defined regions of spacetime~\cite{Morris:1988cz,Visser:1995cc,Lobo:2017cay}.

The first class involves matching an interior wormhole geometry to an exterior Schwarzschild vacuum spacetime across a timelike hypersurface located outside the black hole event horizon. This construction confines the exotic matter required to sustain the wormhole to a finite region bounded by the wormhole throat and the  thin shell at the junction interface. The second class consists of standard thin-shell wormholes formed by the cut-and-paste technique, wherein two identical black hole spacetimes are glued across a shell situated outside their event horizons. These constructions not only offer improved theoretical consistency but also align with the broader goal of formulating semi-classical or quantum gravity-corrected spacetimes that remain regular, geodesically complete, and compatible with asymptotic structures.

This article is organised in the following manner: In Sec. \ref{sec2}, we present the field equations and general properties for traversable wormholes in the string T-duality framework, and find solutions by considering a specific case of a constant redshift function. We also construct a specific model of a wormhole by smoothly joining an interior geometry to an exterior vacuum spacetime. In Sec. \ref{sec3}, we construct a wormhole by applying the cut-and-paste technique, by gluing two identical black hole spacetimes across a timelike hypersurface situated outside their event horizons, and analyse the energy conditions. In Sec. \ref{sec:conclusion}, we summarize and discuss our results, and finally conclude.

\section{Quantum traversable wormholes}\label{sec2}

\subsection{Field equations}

A static, spherically symmetric traversable wormhole can be described by the metric~\cite{Morris:1988cz}
\begin{equation}
	ds^2 = -e^{2\Phi(r)} dt^2 + \frac{dr^2}{1 - b(r)/r} + r^2 (d\theta^2 + \sin^2\theta\, d\phi^2) \,, \label{metricwormhole}
\end{equation}
where $\Phi(r)$ is the redshift function and $b(r)$ is the shape function. The throat of the wormhole occurs at $r = r_0$, where $b(r_0) = r_0$. To be a wormhole solution, the flaring-out condition $(b - b'r)/(2b^2) > 0$~\cite{Morris:1988cz,Visser:1995cc,Lobo:2017cay} must also be satisfied.
To ensure traversability, one requires the absence of horizons, implying that $\Phi(r)$ must remain finite everywhere. Furthermore, the condition $1 - b(r)/r > 0$ must hold throughout the spacetime. To ensure assymptotic flatness, one may also impose that $\Phi \rightarrow 0$ and $b/r \rightarrow 0$ as $r \rightarrow \infty$.

In the present work, we will consistently use the energy density of the modified matter, which is derived from the Poisson equation, as given by Eq. (\ref{eq:rho0}). This approach provides a robust framework for studying the quantum modifications of spacetime geometry, enabling us to explore how quantum effects influence gravitational dynamics at this level. Similarly to how the Poisson equation governs the classical energy distribution, quantum corrections to the Einstein tensor can be obtained by coupling gravity to a quantum-corrected stress-energy tensor, ${\cal T}_{\mu\nu}$, as demonstrated in previous studies \cite{Nicolini:2019irw,Nicolini:2005vd,Modesto:2010uh}.

Assuming natural units ($G = c = 1$), the Einstein field equations, $G_{\mu\nu}= 8 \pi {\cal T}_{\mu\nu}$, yield the following stress-energy components:
\begin{eqnarray}
	\rho(r) &=& \frac{1}{8\pi} \frac{b'(r)}{r^2} \,, 
	\label{rhoWH}
		\\
	p_r(r) &=& \frac{1}{8\pi} \left[ -\frac{b(r)}{r^3} + 2\left(1 - \frac{b(r)}{r} \right)\frac{\Phi'(r)}{r} \right] \,, 
	\label{prWH} 
		\\
	p_t(r) &=& \frac{1}{8\pi} \left(1 - \frac{b(r)}{r}\right) \Bigg[ \Phi'' + (\Phi')^2 + \frac{\Phi'}{r}  
		\nonumber \\
	&& - \frac{b'r - b}{2r(r - b)}\Phi' - \frac{b'r - b}{2r^2(r - b)} \Bigg] \,. \label{ptWH}
\end{eqnarray}
Here, $\rho(r)$ is the energy density, $p_r(r)$ is the radial pressure, and $p_t(r)$ is the transverse pressure, respectively.

The conservation of the stress-energy tensor, $T^{\mu\nu}{}_{;\nu} = 0$, imposes the constraint:
\begin{equation}
	p_r'(r) = \frac{2}{r} (p_t - p_r) - (\rho + p_r)\Phi'(r) \,, \label{prderivative}
\end{equation}
which may be interpreted as the anisotropic Tolman-Oppenheimer-Volkov equation.
The term in the left-hand-side can be considered as the hydrostatic force, the first term in the right-hand-side as an anisotropic force and the second term is the relativistic gravitational force.
One may define the anisotropic factor as $\Delta = p_t -p_r$, which measures the pressure anisotropy of the fluid comprising the wormhole, and acts as an effective additional force. For a positive anisotropy $p_t >p_r$, there is an effective force pushing outward, which supports equilibrium. For a negative anisotropy $p_t < p_r$, there is an inward attractive pull, implying collapse.

A defining feature of traversable wormholes is the violation of the null energy condition (NEC), defined via $T_{\mu\nu}k^\mu k^\nu \geq 0$ for any null vector $k^\mu$. The NEC is given by
\begin{equation}
	T_{\mu\nu}k^\mu k^\nu = \rho(r) + p_r(r) = \frac{1}{8\pi} \left[\frac{b'r - b}{r^3} + 2\left(1 - \frac{b}{r}\right)\frac{\Phi'}{r} \right] \,. \label{NECthroat}
\end{equation}
Evaluating this expression at the throat and using the flaring-out condition $(b - b'r)/(2b^2) > 0$, mentioned above, together with the regularity of $\Phi(r)$, confirms the violation of the NEC. Matter that violates this condition is typically referred to as \textit{exotic}~\cite{Morris:1988cz}.
Additionally, the flaring-out condition requires $b'(r_0) < 1$ at the throat.

To construct explicit wormhole solutions, one may employ a variety of approaches, depending on the physical and mathematical constraints of the system. It is important to recognize that the governing system of equations is inherently underdetermined, comprising three equations, namely, Eqs.~(\ref{rhoWH})--(\ref{ptWH}), which relate five unknown functions: the energy density $\rho(r)$, radial pressure $p_r(r)$, tangential pressure $p_t(r)$, redshift function $\Phi(r)$, and shape function $b(r)$.
However, since the energy density has been specified a priori, as given by Eq.~(\ref{eq:rho0}), this effectively reduces the number of free functions, thereby providing an additional degree of freedom to constrain the system. To proceed analytically, and for the sake of simplicity, we shall adopt a particular ansatz for the redshift function $\Phi(r)$ in what follows.

\subsection{String T-duality-inspired wormhole}\label{sec:BH}

Rewriting the energy density (\ref{eq:rho0}) considered in the Introduction, i.e.,
\begin{equation}
	\rho(r)
	= \frac{3 l_0^2 M}{4 \pi {\left(r^2 +l_0^2\right)}^{5/2}}\,,
	\label{eq:rho}
\end{equation}
one verifies that for large distances, the above density quickly falls off as $~r^{-5}$, and at short scales $r\lesssim l_0$, one encounters a regular quantum region characterized by the creation and annihilation of virtual particles at a constant and finite energy. Here, gravity becomes repulsive and prevents the full collapse of matter into a singularity \cite{Nicolini:2023hub}.

Thus, substituting Eq. (\ref{eq:rho}) into Eq.~(\ref{rhoWH}), the integration yields the following shape function:
\begin{equation}
	b(r) = r_0 + \frac{2 r^3 M}{ {\left(r^2 +l_0^2\right)}^{3/2}} 
	- \frac{2 r_0^3 M}{ {\left(r_0^2 +l_0^2\right)}^{3/2}}\,,
	\label{eq:shape}
\end{equation}
where the condition $b(r_0)=r_0$ has been imposed to define explicitly the constant of integration.

\begin{figure}[t]
	\includegraphics[width=\linewidth]{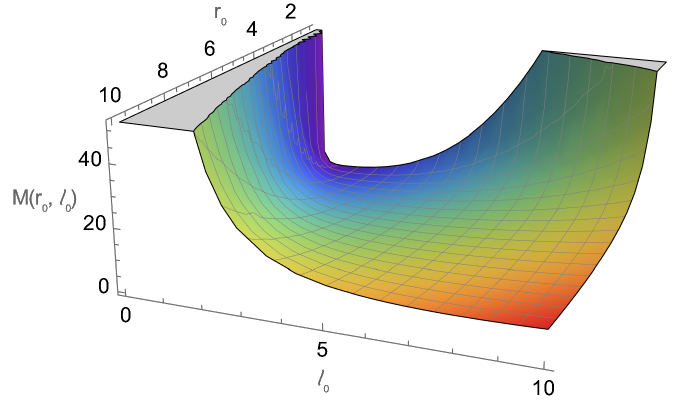}
	\caption{The surface depicts the function $M = {\left(r_0^2 +l_0^2\right)}^{5/2}/( 6 r_0^2 l_0^2)$. The parameter range is only valid below the surface, which satisfies the flaring-out condition, given by the inequality \ref{eq:dshape}. Qualitatively, this implies that for low values of $l_0$ and large values of $r_0$, one may obtrain large values for $M$.}
	\label{fig1_flareout}
\end{figure}

Note that the flaring-out condition, at the throat, imposes the following condition 
\begin{equation}
	b'(r_0) = \frac{6 r_0^2 l_0^2 M}{ {\left(r_0^2 +l_0^2\right)}^{5/2}}<1 \,,
	\label{eq:dshape}
\end{equation}
which provides a constraint on the parameters of the model. Thus, this can be cast into the following form
\begin{equation}
	M < \frac{{\left(r_0^2 +l_0^2\right)}^{5/2}}{ 6 r_0^2 l_0^2}\,,
\end{equation}
which is depicted in Fig. (\ref{fig1_flareout}).
For $r_0 \approx l_0 $, we verify that $M < 2\sqrt{2}l_0/3$; for $r_0 \gg l_0 $, then $M < r_0^3/(6l_0^2)$, where the mass may assume arbitrary large values.

Note that the inequality (\ref{eq:dshape}) implies that the NEC is violated at the throat, as expected:
\begin{equation}
	\left[\rho(r) +  p_r(r)\right]\Big|_{r_0} = - \frac{\left(r_0^2 +l_0^2\right)^{5/2}-6 l_0^2 r_0^2 M}{ 8\pi r_0^2
		{\left(r_0^2 +l_0^2\right)}^{5/2}} <0\,.
	\label{eq:shape2}
\end{equation}

\subsection{Constant redshift function: $\Phi'(r)=0$}

One may now consider several strategies to obtain wormhole solutions, however, for simplicity we consider the specific case of a constant redshift function, i.e., with $\Phi'(r)=0$. In addition to the string T-duality-inspired energy density (\ref{eq:rho}), the radial and tangential pressures are given by:
\begin{equation}
	p_r\left(r \right) = 
	-\frac{1}{8 \pi r^{3}}
	\left[r_0 +\frac{2 r^{3} M}{\left(l_0^{2}+r^{2}\right)^{3/2}}-\frac{2 r_0^{3} M}{\left(r_0^{2}+l_0^{2}\right)^{3/2}}\right]\,,
\end{equation}
\begin{eqnarray}
	p_t(r)&=&
	- \frac{1}{16 \pi  r^{3}}
	\Bigg[\left(\frac{6 r^{2} M}{\left(l_0^{2}+r^{2}\right)^{3/2}}-\frac{6 r^{4} M}{\left(l_0^{2}+r^{2}\right)^{5/2}}\right) r 
		\nonumber \\
	&& -r_0 -\frac{2 r^{3} M}{\left(l_0^{2}+r^{2}\right)^{3/2}}+\frac{2 r_0^{3} M}{\left(r_0^{2}+l_0^{2}\right)^{3/2}}\Bigg]\,,
\end{eqnarray}
respectively. 

As the redshift function is zero, the relativistic gravitational force that appears in the right-hand-side of Eq. (\ref{prderivative}) is zero. The term $dp_r/dr$ in the left-hand-side is given by
\begin{eqnarray}
	\frac{dp_r}{dr} &=& \frac{3}{8\pi r^{3}} \Bigg[\frac{r_0}{r}+\frac{2 r^{2} M}{\left(l_0^{2}+r^{2}\right)^{3/2}}-\frac{2 r_0^{3} M}{r\left(r_0^{2}+l_0^{2}\right)^{3/2}}
	\nonumber \\
	&&-\frac{r^{2} M}{\left(l^{2}+r^{2}\right)^{3/2}}+\frac{r^{4} M}{\left(l^{2}+r^{2}\right)^{5/2}}\Bigg] \,.
	\label{dp_over_dr}
\end{eqnarray}
A trivial, but lengthy, calculation yields that the anistropic pressure force $2\Delta/r$ is equal to Eq. (\ref{dp_over_dr}), for self-consistency.

The NEC is provided by the following relation:
\begin{eqnarray}
	{p_r}(r) + \rho(r) &= &
	-	\frac{1}{8 \pi  \,r^{3}}
	\Bigg[r_0 +\frac{2 r^{3} M}{\left(l_0^{2}+r^{2}\right)^{3/2}}
	\nonumber\\
	&&
	-\frac{2 r_0^{3} M}{\left(r_0^{2}+l_0^{2}\right)^{3/2}}
		 +\frac{6 l_0^{2} M r^3}{ \left(l_0^{2}+r^{2}\right)^{5/2}}\Bigg]\,,
\end{eqnarray}
where it is straightforward to show that the NEC is violated throughout the spacetime geometry.

An alternative approach involves considering a specific shape function and subsequently solving the Einstein field equations for the redshift function. However, this typically results in exceedingly lengthy and cumbersome expressions. Moreover, these solutions tend to exhibit violations of the NEC across the entire spacetime manifold. Given that these configurations closely resemble the ones discussed above, we do not reproduce them here. Instead, we are motivated to explore scenarios in which the NEC violations are minimized or localized. To this end, we consider the possibility of matching the interior wormhole geometry to an exterior vacuum solution—such as the Schwarzschild spacetime—which can constrain the region of the NEC violation. This strategy is explored in the following subsection.

\subsection{Wormhole surrounded by a thin shell}

In this subsection, we construct a specific model of a wormhole by smoothly matching an interior geometry to an exterior vacuum spacetime. The matching is performed across a timelike hypersurface $\partial \Omega$, defined by the fixed radial coordinate $r = a$, which serves as the junction interface between the two regions. For simplicity, we adopt the Schwarzschild metric as the exterior geometry and choose the matching radius such that $a > r_b = 2M$, where $r_b$ is the Schwarzschild radius. This condition ensures that the hypersurface $\partial \Omega$ lies outside the event horizon, thereby maintaining the traversable nature of the wormhole \cite{Visser:1995cc,Garcia:2011aa,Lobo:2005zu,Lobo:2004id,Lobo:2004rp,Sushkov:2005kj,Lobo:2005us,Lobo:2005yv,Lobo:2005vc,Lobo:2006ue}.

To analyze the matching conditions at the interface $\partial \Omega$, we employ the Darmois-Israel formalism \cite{Darmois,Israel:1966rt}, which yields the surface stress-energy tensor $S^i_{\;j}$ via the Lanczos equations:
\begin{equation}
	S^i_{\;j} = -\frac{1}{8\pi} \left( \kappa^i_{\;j} - \delta^i_{\;j} \kappa^k_{\;k} \right) \,,
	\label{eq:Lanczos}
\end{equation}
where the quantity $\kappa_{ij}$ denotes the jump in the extrinsic curvature across the junction surface, defined as $\kappa_{ij} = K_{ij}^{+} - K_{ij}^{-}$. The extrinsic curvature itself is expressed as $K_{ij} = n_{\mu;\nu} \, e^{\mu}_{(i)} e^{\nu}_{(j)} $ with $n^{\mu}$ representing the unit normal vector to $\partial \Omega$ and $e^{\mu}_{(i)}$ denoting the basis vectors tangent to the hypersurface.

For the specific configuration under consideration, namely, the junction between an interior wormhole geometry, given by Eqs. \ref{metricwormhole}, and a Schwarzschild exterior, the non-vanishing components of the extrinsic curvature on both sides of the shell are given by:
\begin{eqnarray}
	K^{\tau +}_{\;\;\tau} &=& \frac{\frac{M}{a^2}}{\sqrt{1 - \frac{2M}{a}}} \,, \label{eq:KtautauPlus} \\
	K^{\tau -}_{\;\;\tau} &=& \Phi'(a) \,\sqrt{1-\frac{b(a)}{a}} \,, \label{eq:KtautauMinus}
\end{eqnarray}
and
\begin{align}
	K^{\theta +}_{\;\;\theta} &= \frac{1}{a} \sqrt{1 - \frac{2M}{a}} \,, \label{eq:KthetaPlus} \\
	K^{\theta -}_{\;\;\theta} &= \frac{1}{a} \sqrt{1 - \frac{b(a)}{a}}\,, 
	\label{eq:KthetaMinus}
\end{align}
respectively.

Substituting these into Eq.~(\ref{eq:Lanczos}) allows one to determine the explicit expressions for the surface stresses at the junction, given by:
\begin{eqnarray}
	\sigma&=&-\frac{1}{4\pi a} \left(\sqrt{1-\frac{2M}{a}}-
	\sqrt{1-\frac{b(a)}{a}} \, \right)
	\label{surfenergy}   ,\\
	{\cal P}&=&\frac{1}{8\pi a} \left[\frac{1-\frac{M}{a}}{\sqrt{1-\frac{2M}{a}}}
	-(1+a\Phi') \sqrt{1-\frac{b(a)}{a}}
	\, \right]  ,   
	\label{surfpressure}
\end{eqnarray}
where $\sigma$ and ${\cal P}$ are the surface energy density and
the tangential surface pressure, respectively.
Note that if the surface stress-energy terms are zero, the junction is
denoted as a boundary surface. If surface stress terms are
present, the junction is called a thin shell. The
surface mass of the thin shell is given by $m_s=4\pi a^2\sigma$.

Consider now the NEC at the thin shell, i.e., $\sigma + {\cal P}$.
For simplicity, assume the specific case of a constant redshift function $\Phi'=0$, and taking into account the shape function (\ref{eq:shape}), then Eqs. (\ref{surfenergy}) and (\ref{surfpressure}) yield:
\begin{eqnarray}
	\sigma + {\cal P} &=& \frac{1}{8\pi a}	\Bigg[\sqrt{1-\frac{r_0}{a}+\frac{2 r_0^{3} M}{a\left(r_0^{2}+l_0^{2}\right)^{3/2}}-\frac{2 a^{2} M}{\left(a^{2}+l_0^{2}\right)^{3/2}}}
	\nonumber \\
	&& \qquad \qquad - \frac{1-\frac{3M}{a}}{\sqrt{1-\frac{2 M}{a}}}\; \Bigg]\,.
\end{eqnarray}

In summary, the strategy of matching an interior wormhole geometry to an exterior vacuum spacetime, such as the Schwarzschild solution, offers a physically motivated means of minimizing the violation of the NEC. By confining the exotic matter content to within a thin shell at the junction interface and the wormhole throat $r_0$, this construction localizes the NEC violations to a finite region. Moreover, situating the matching surface outside the event horizon ensures the traversability of the configuration while maintaining compatibility with an asymptotically flat external geometry.

\section{Quantum-inspired thin-shell wormholes}\label{sec3}

\subsection{Cut-and-paste procedure}

In this section, we construct a wormhole solution by employing the well-known cut-and-paste technique \cite{Visser:1995cc}. The procedure begins by considering two identical copies of a static black hole spacetime, from each of which we excise the region defined by
\[
\Omega_{1,2} \equiv \left\{ r_{1,2} \leq a \,\middle|\, a > r_b \right\},
\]
where \(a\) is a constant denoting the radius of the junction interface and \(r_b\) represents the radius of the event horizon of the black hole. This excision removes the interior region containing the singularity and the horizon.

The removal of these regions yields two geodesically incomplete manifolds, each possessing a boundary described by the timelike hypersurfaces
\[
\partial \Omega_{1,2} \equiv \left\{ r_{1,2} = a \,\middle|\, a > r_b \right\}.
\]
By identifying these two hypersurfaces via the condition \(\partial \Omega_{1} = \partial \Omega_{2}\), we obtain a new geodesically complete manifold in which the two asymptotically flat regions are connected through a throat located at the junction surface \(\partial \Omega\) \cite{Poisson:1995sv,Visser:1995cc,Lobo:2003xd,Eiroa:2003wp,Ishak:2001az,Lemos:2008aj,Lemos:2004vs}.

To analyze the physical properties of the throat, particularly the stress-energy content required to support it, as before, we invoke the Darmois-Israel formalism \cite{Darmois,Israel:1966rt}. Recall that this formalism allows for a rigorous treatment of thin-shell spacetimes by characterizing the surface stress-energy tensor \(S_{ij}\) residing on the junction hypersurface, which is given by Eq. (\ref{eq:Lanczos}).

\subsection{Lanczos equations and energy conditions}

We consider, in particular, the matching of two static and
spherically symmetric spacetimes given by the following line
elements
\begin{equation}
	ds^2=-e^{2\alpha(r)}\,dt^2
	+e^{-2\alpha(r)}\,dr^2+r^2(d\theta
	^2+\sin ^2{\theta}\, d\phi ^2)  \,,
	\label{generalmetric}
\end{equation}
where the metric function is given by 
\begin{equation}
	e^{2\alpha(r)}  = 1-\frac{2Mr^2}{(r^2+l_0^2)^{3/2}}\,.
	\label{Def:BH}
\end{equation}
In order to guarantee a static spacetime, the condition $e^{2\alpha(r)} > 0 $ imposes that 
\begin{equation}
	M < \frac{(a^2+l_0^2)^{3/2}}{2a^2}\,.
	\label{conditionMshell}
\end{equation}

The non-trivial components of the extrinsic curvature are given by
\begin{eqnarray}
	K ^{\theta
		\;\pm}_{\;\;\theta}&=& \pm \frac{1}{a}\,e^{\alpha}
		= \pm \frac{1}{a} \sqrt{1-\frac{2Mr^2}{(r^2+l_0^2)^{3/2}}} \;,
	\label{genKplustheta}\\
	K ^{\tau \;\pm}_{\;\;\tau}&=& \pm \alpha' e^{\alpha}
		= \pm \frac{Ma (a^2-2l_0^2)}{(a^2+l_0^2)^{5/2}\sqrt{1-\frac{2Mr^2}{(r^2+l_0^2)^{3/2}}}}
	\;. \label{genKminustautau}
\end{eqnarray}
The Lanczos equations (\ref{eq:Lanczos}), then provide us with the following expressions for the surface stresses
\begin{eqnarray}
	\sigma&=&-\frac{1}{2\pi a} e^{\alpha}
	\label{gen-surfenergy2}   ,\\
	{\cal P}&=&\frac{1}{4\pi a} \left(1+a\alpha'\right)
		e^{\alpha}
	\label{gen-surfpressure2}    \,,
\end{eqnarray}
and using the metric function (\ref{Def:BH}), these take the following form:
\begin{eqnarray}
	\sigma&=&-\frac{1}{2\pi a} \, \sqrt{1-\frac{2Ma^2}{(a^2+l_0^2)^{3/2}}}
	\label{gen-surfenergyBH}   \,,\\
	{\cal P}&=&\frac{1}{4\pi a} \left[1+ \frac{a\left(\frac{3Ma^3}{\left( a^2 + l_0^2 \right)^{5/2}} - \frac{2Ma}{\left( a^2 + l_0^2 \right)^{3/2}} \right)}{1-\frac{2Ma^2}{(a^2+l_0^2)^{3/2}}} \right]\times
		\nonumber \\
	&& \qquad \qquad\times \sqrt{1-\frac{2Ma^2}{(a^2+l_0^2)^{3/2}}}
	\label{gen-surfpressureBH}    \,.
\end{eqnarray}
The surface energy density in (\ref{gen-surfenergyBH}) is negative, and thus 
implies the violation of the weak and dominant energy conditions \cite{Visser:1995cc}.

Consider the NEC at the junction interface:
\begin{equation}
\sigma + {\cal P} = \frac{1}{4\pi a} \frac{\left[3 M a^{4}-\left(a^{2}+l_0^{2}\right)^{5/2}\right]}{\left(a^{2}+l_0^{2}\right)^{5/2}\sqrt{1-\frac{2 M a^{2}}{\left(a^{2}+l_0^{2}\right)^{3/2}}}} \,.
\end{equation}
In order to have the non-violation of the NEC, i.e., $\sigma + {\cal P} \geq 0$, and taking into account the condition (\ref{conditionMshell}), one verifies that $a > \sqrt{2} l_0$ is imposed, which is a physically reasonable assumption.

The strong energy condition (SEC) is satisfied
if $\sigma+{\cal P} >0$ and $\sigma+2{\cal P} >0$, and by
continuity implies the NEC. Using the condition
\begin{eqnarray}
	\sigma+2{\cal P}&=&-\frac{1}{2\pi}\frac{a M \left(a^{2}-2 l^{2}\right)}{\left(a^{2}+l^{2}\right)^{5/2}  \sqrt{1-\frac{2 M \,a^{2}}{\left(a^{2}+l^{2}\right)^{\frac{3}{2}}}}}
	\,,
	\label{sigma+2P}
\end{eqnarray}
we verify that $\sigma+2{\cal P} < 0$ for $a > \sqrt{2} l_0$, and thus the SEC is violated in this region.

\section{Summary and Discussion}\label{sec:conclusion}

In this work, we have investigated the construction of spherically symmetric, electrically neutral wormhole geometries within a framework inspired by an ultraviolet (UV) completion of gravity, particularly those rooted in string-theoretic dualities. By taking into account the correspondence between T-duality and path integral duality, we introduced a fundamental minimal length scale—identified as the zero-point length—that effectively regularizes the short-distance behavior. This scale encapsulates non-perturbative quantum gravitational effects and serves as a natural regulator, ensuring that the resulting spacetimes are geodesically complete and free from curvature singularities.

In the context of  wormhole physics, a persistent challenge is the inevitable violation of the null energy condition (NEC), typically interpreted as the requirement of exotic matter to support the wormhole throat. Our analysis confirms that solutions based on constant redshift functions, or those constructed by prescribing specific shape functions, generally lead to NEC violations distributed throughout the entire spacetime. While such configurations are analytically tractable, they often result in highly intricate expressions and extended regions of exotic matter, reducing their physical plausibility.

To address this limitation, we focused on wormhole models in which the violation of the NEC is spatially localized and minimized. We explored two classes of thin shell constructions that achieve this goal. The first involves matching a regular interior wormhole geometry to an exterior Schwarzschild vacuum across a timelike hypersurface situated outside the black hole event horizon. In this scenario, the exotic matter required to sustain the throat is confined to a finite, localized region within the junction interface. The second approach uses the standard cut-and-paste technique, wherein two identical copies of a regular black hole spacetime are joined across a timelike surface. In both cases, the resulting wormhole geometries restrict the NEC violations to compact regions.

Our results demonstrate that ultraviolet modifications inspired by string theory, particularly those induced by T-duality, can be consistently implemented within semiclassical gravitational frameworks to yield physically admissible wormhole solutions. These spacetimes not only resolve the classical curvature singularities but also maintain asymptotic flatness, geodesic completeness, and traversability. As such, they represent promising candidates for quantum-corrected wormhole geometries that are compatible with known physical principles.

Looking forward, several extensions of this work merit further investigation. These include generalizations to rotating or time-dependent wormhole solutions, the inclusion of electromagnetic charges or a cosmological constant, and a detailed analysis of the dynamical stability of these configurations under perturbations. Additionally, a more comprehensive exploration of the connection between T-duality-induced corrections and energy condition violations could offer deeper insights into the nature of exotic matter and the quantum microstructure of spacetime near the Planck scale.

\acknowledgments{FSNL acknowledges support from the Funda\c{c}\~{a}o para a Ci\^{e}ncia e a Tecnologia (FCT) Scientific Employment Stimulus contract with reference CEECINST/00032/2018, and funding through the research grants UIDB/04434/2020, UIDP/04434/2020 and PTDC/FIS-AST/0054/2021.
MER thanks Conselho Nacional de Desenvolvimento Cient\'{\i}fico e Tecnol\'ogico - CNPq, Brazil, for partial financial support.}

\end{document}